\documentclass[twocolumn,prl,aps,superscriptaddress,footinbib,amsfonts,amsmath,amssymb,showpacs,floatfix]{revtex4-2}
\usepackage[utf8]{inputenc}
\usepackage{multirow}
\usepackage{graphicx}
\usepackage{dsfont}
\usepackage{amsfonts}
\usepackage{amsmath}
\usepackage{amsthm}
\usepackage{bbm,bbold}
\usepackage{float}
\usepackage[normalem]{ulem}
\usepackage[dvipsnames]{xcolor}
\definecolor{DartmouthGreen}{RGB}{0, 105, 62}
\usepackage{physics}
\usepackage[utf8]{inputenc}

\usepackage{mathtools}
\usepackage{amssymb}
\usepackage{tikz}
\usetikzlibrary{quantikz}
\usetikzlibrary{shapes,snakes}

\usepackage{pgfplots}
\pgfplotsset{compat=1.13}

 \usepackage{graphicx}
 \usepackage[colorlinks=true, linkcolor=green, citecolor=blue]{hyperref}

 \def\be{\begin{equation}}
 \def\ee{\end{equation}}
 \def\bea{\begin{eqnarray}}
 \def\eea{\end{eqnarray}}
 \def\bean{\begin{eqnarray*}}
 \def\eean{\end{eqnarray*}}
 \def\gsim{\mathrel{\rlap{\lower0.2em\hbox{$\sim$}}\raise0.2em\hbox{$>$}}}
 \def\ksim{\mathrel{\rlap{\lower0.2em\hbox{$\sim$}}\raise0.2em\hbox{$<$}}}
 \def\kg{\mathrel{\rlap{\lower0.25em\hbox{$>$}}\raise0.25em\hbox{$<$}}}

\newtheorem{theorem}{Theorem}

\begin{document}
\title{Reducing the qubit requirement of Jordan-Wigner encodings\\ of $N$-mode, $K$-fermion systems from $N$ to $\lceil \log_2 {N \choose K} \rceil$}



\author{Brent Harrison} 
\email[Email: ]{Brent.A.Harrison.GR@dartmouth.edu}
\affiliation{Department of Physics and Astronomy, Dartmouth College, Hanover, New Hampshire 03755, USA}
\author{Dylan Nelson}
\email[Email: ]{dylan.nelson@alumni.uct.ac.za}
\noaffiliation{}

\author{Daniel Adamiak }
\email[Email: ]{adamiak.5@osu.edu}
\affiliation{Department of Physics, The Ohio State University, Columbus, Ohio 43210, USA}

\author{James Daniel Whitfield}
\email[Email: ]{James.D.Whitfield@Dartmouth.edu}
\affiliation{Department of Physics and Astronomy, Dartmouth College, Hanover, New Hampshire 03755, USA}
\affiliation{AWS Center for Quantum Computing, Pasadena, California 91106, USA}

\begin{abstract}
    To simulate a fermionic system on a quantum computer, it is necessary to encode the state of the fermions onto qubits. Fermion-to-qubit mappings such as the Jordan-Wigner and Bravyi-Kitaev transformations do this using $N$ qubits to represent systems of $N$ fermionic modes. In this work, we demonstrate that for particle number conserving systems of $K$ fermions and $N$ modes, the qubit requirement can be reduced to the information theoretic minimum of $\lceil \log_2 {N \choose K} \rceil$. This will improve the feasibility of simulation of molecules and many-body systems on near-term quantum computers with limited qubit number. 
\end{abstract}

\maketitle





\section{Introduction}




The simulation of interacting fermionic systems is computationally hard on a classical computer, but in principle tractable on quantum hardware~\cite{brod_bosons_2021, brown_using_2010}. This important application of quantum computers, conceived of by Feynman in 1982~\cite{feynman_simulating_1982} and further described by Lloyd in 1996~\cite{lloyd_universal_1996},
has relevance to diverse problems in quantum chemistry, many-body physics and material science.

Since qubits do not respect fermionic antisymmetry by default, any simulation of fermions on a quantum computer necessarily entails the construction of encoded qubit representations of fermionic states and operators~\cite{Bravyi_2002}. There are several known fermion-to-qubit encodings; particular examples include the Jordan-Wigner~\cite{Jordan:1928wi}, parity basis~\cite{doi:10.1063/1.4768229} and Bravyi-Kitaev~\cite{Bravyi_2002, doi:10.1063/1.4768229} mappings.

These encodings can be used to map a fermionic Hamiltonian to a qubit Hamiltonian $H = \sum_i H_i$, where the $H_i$ are tensor products of Pauli operators (``Pauli strings''). The time evolution of this Hamiltonian can then be simulated by a quantum computer, e.g.~by making use of the Trotter expansion~\cite{lloyd_universal_1996} to divide the propagator $U = \exp(-iHt)$ into a product of short evolutions generated by the terms $H_i$. These short evolutions are then further decomposed into gate sequences.

While many choices of fermion-to-qubit encodings exist, it is desirable to construct encodings that minimize the use of resources such as circuit depth and qubit requirement. The former increases with the operator locality~\cite{havlicek_operator_2017} or ``Pauli weight'' of each encoded term $H_i$, i.e. the number of qubits the term acts on non-trivially. The latter can be reduced by taking advantage of symmetries of the system.

For example, the Jordan-Wigner and parity basis encodings both  require $N$ qubits to encode $N$ fermionic modes. However, the latter can be implemented with just $N-1$ qubits in the presence of parity conservation. In information theoretic terms, the known overall parity provides one bit of information entropy. It seems then that in the presence of particle number conservation, it should be possible to extract more than one bit of information entropy and potentially reduce the qubit requirement still further.
 

\begin{figure*}
    \centering
    \includegraphics[width=\linewidth]{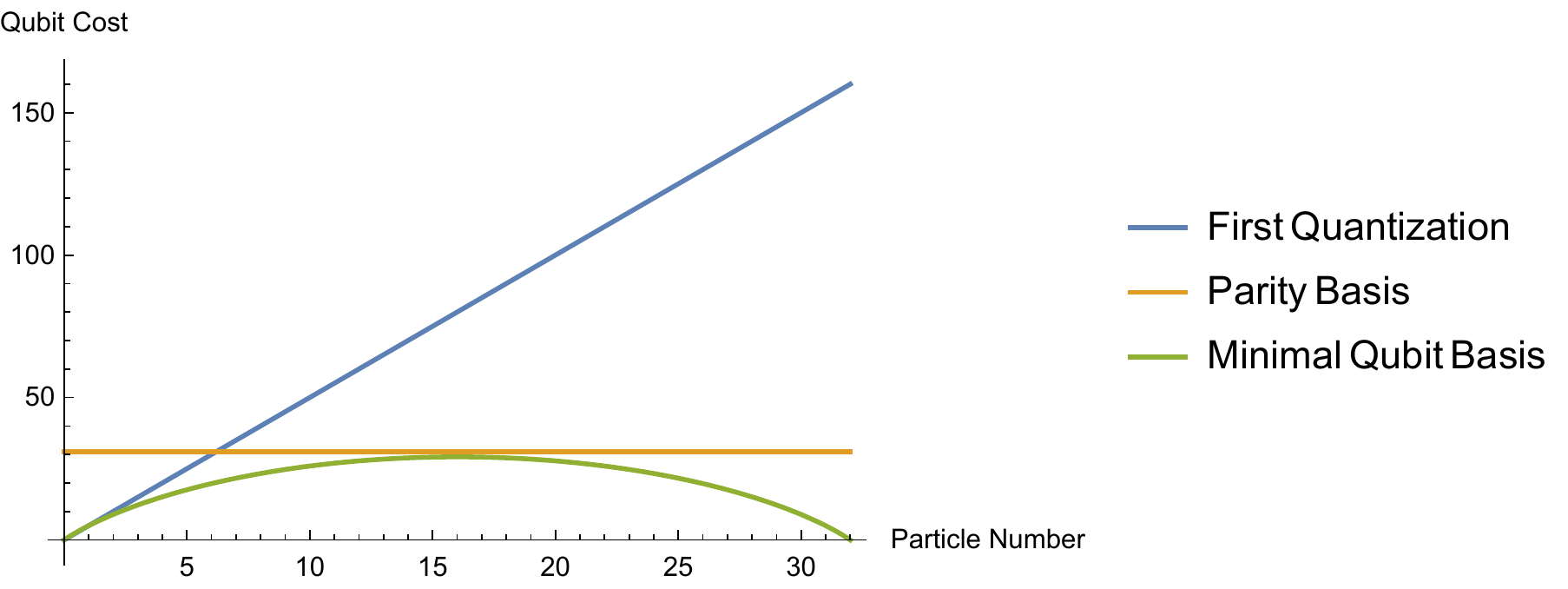}
    \caption{Qubit cost of representing a  $32$-mode system with varying number of fermions using first quantization (scaling as $K\log_2 N$~\cite{kassal2008polynomial,zalka1998simulating}), the parity mapping with known overall parity ($N - 1$), and the minimal qubit basis presented here ($\lceil \log_2 {N \choose K} \rceil$).}
    \label{fig:qubitReductionIdea}
\end{figure*}

In this paper, we will show that for particle-number conserving systems of $K$ fermions occupying $N$ modes, it is possible to construct a class of fermion-to-qubit encodings analogous to Jordan-Wigner, but with a qubit requirement reduced to the information theoretic minimum~\cite{Kirby:2021vkt} of $\lceil \log_2 {N \choose K} \rceil$.  

The outline of the paper is as follows. We begin by introducing the Jordan-Wigner and parity basis mappings. We then show how to construct a mapping from the Jordan-Wigner basis to the parity basis. Since each basis contains the same elements, this mapping is just a permutation of those elements.
We do this in anticipation of the next section, in which we show how to construct similar permutations mapping to bases in which the qubit requirement can be reduced to $\lceil \log_2 {N \choose K} \rceil$. See Fig.~\ref{fig:qubitReductionIdea}.

We then discuss how choosing a permutation that is also an element of the Clifford group reduces the circuit depth of the encoding, but show that such Clifford permutations can reduce the qubit requirement by at most one qubit.

We explicitly construct an example minimal qubit basis for a 4-mode system. We then compare our results to other qubit cost reduction schemes for fermion-to-qubit encodings in second quantization. Finally we conclude and present an outlook for future work. 

\section{The Jordan-Wigner and Parity Basis Encodings}\label{sec:JWParity}

\subsection{Jordan-Wigner}

In general, the state of a second-quantized fermionic system can be written in terms of a binary string,
\begin{equation}
\begin{aligned}
    \ket{\psi} &= (a^\dagger_1)^{n_1}(a^\dagger_2)^{n_2}\dots(a^\dagger_N)^{n_N}\ket{00\dots 0}\\
    &= \ket{n_1 n_2 \dots n_N}.
\end{aligned}
\end{equation}
Here $a^\dagger_j$ and $a_j$ are the fermionic creation and annihilation operators for mode $j$, $n_j \in \{0,1\}$ is the occupancy number of mode $j$, and the state $\ket{00\dots 0}$ is the vacuum.

Then in particular, the state $\ket{10011}$ represents 3 fermions occupying modes 1, 4 and 5 of a 5-mode system. The most obvious way to encode such a state onto qubits is to directly map this binary string onto an identical binary string stored on a qubit register,
\begin{equation}\label{eq:JWStateMapExample}
    \ket{10011}_{\text{fermions}} \rightarrow \ket{10011}_{\text{qubits}} .
\end{equation}

It is then necessary to construct appropriate encoded qubit creation and annihilation operators that preserve the fermionic anticommutation relations,
\begin{equation}
\{a_j, a_k^\dagger\} = \delta_{jk}I, \hspace{2mm} \{a_j,a_k\} = \{a_j^\dagger, a_k^\dagger\} = 0.
\end{equation}
These are given by the well-known Jordan-Wigner transform~\cite{Jordan:1928wi},
\begin{equation}
\begin{aligned}
    a^\dagger_j \rightarrow \frac{1}{2}\prod_{k<j} Z_k(X_j - iY_j),\\
    a_j \rightarrow \frac{1}{2}\prod_{k<j} Z_k(X_j + iY_j).
\end{aligned}
\end{equation}

It will be convenient to work in the Majorana basis,
\begin{equation}
\begin{aligned}
    \gamma_j \equiv a^\dagger_j + a_j, \hspace{2mm} \gamma_j' \equiv i(a^\dagger_j - a_j).
\end{aligned}
\end{equation}
Note that all Majoranas square to the identity and mutually anticommute,
\begin{equation}\label{eq:MajoranaCommutationRelations}
    \{\gamma_j,\gamma_k\} = \{\gamma_j',\gamma_k'\} = 2\delta_{jk}I, \hspace{2mm} \{\gamma_j,\gamma_k'\} = 0.
\end{equation}

Then under the Jordan-Wigner mapping,
\begin{equation}
    \gamma_j \rightarrow \prod_{k<j} Z_kX_j, \hspace{3mm}
    \gamma_j' \rightarrow \prod_{k<j} Z_kY_j.
\end{equation}

\subsection{The Parity Basis}

Now, the straightforward mapping \eqref{eq:JWStateMapExample} is by no means the only way to represent the state of a fermionic system on qubits. An equally valid mapping can be constructed in terms of parity sums,
\begin{equation}
    \ket{n_1n_2\dots n_N}_{\text{fermions}} \rightarrow \ket{x_1x_2\dots x_N}_{\text{qubits}},
\end{equation}
where $x_j \equiv n_1 \oplus n_2 \oplus \dots \oplus n_j$, and $\oplus$ signifies addition modulo 2. Our example state then gets mapped to qubits by
\begin{equation}
    \ket{10011}_{\text{fermions}} \rightarrow \ket{11101}_{\text{qubits}}.
\end{equation}

The corresponding creation and annihilation operators are~\cite{doi:10.1063/1.4768229}
\begin{equation}
\begin{aligned}
    a_j^\dagger &\rightarrow \frac{1}{2}\left(Z_{j-1}X_j - iY_j\right)\prod_{k>j} X_k,\\
    a_j &\rightarrow \frac{1}{2}\left(Z_{j-1}X_j + iY_j\right)\prod_{k>j} X_k,
\end{aligned}
\end{equation}
with Majoranas given by
\begin{equation}
\begin{aligned}
    \gamma_j \rightarrow Z_{j-1}X_j\prod_{k>j} X_k, \hspace{3mm}
    \gamma_j' \rightarrow Y_j\prod_{k>j} X_k.
\end{aligned}
\end{equation}

The parity basis has the advantage that a symmetry of the system is made manifest in the qubit representation. If the overall parity of the system is fixed, then the final qubit always has the same value, and is thus redundant. The parity basis mapping therefore allows us to represent fermionic states with one qubit fewer than Jordan-Wigner.

As we will see, it is possible to find a basis that reduces the qubit requirement still further by taking advantage of the information entropy associated with particle number conservation. We will do this by constructing a mapping from Jordan-Wigner to the desired basis. To introduce this concept, we will first describe how to construct a mapping from Jordan-Wigner to the parity basis.

\section{Basis Permutations}\label{sec:basisPermutations}

We can relate the Jordan-Wigner transform to the parity basis via a permutation of the $2^N$-dimensional computational basis. This is easy to see for a 2-mode system; consider an operator $P_{JW\rightarrow P}$ that acts on the basis  $\{\ket{00},\ket{01},\ket{10},\ket{11}\}$ as
\begin{equation}
\begin{aligned}
P_{JW\rightarrow P} \ket{00} &= \ket{00}, \hspace{3mm} P_{JW\rightarrow P} \ket{01} = \ket{01},\\
P_{JW\rightarrow P} \ket{10} &= \ket{11}, \hspace{3mm} P_{JW\rightarrow P} \ket{11} = \ket{10}.
\end{aligned}
\end{equation}
Then the matrix representation of $P_{JW\rightarrow P}$ is
\begin{equation}
P_{JW\rightarrow P} = \begin{bmatrix}
1 & 0 & 0 & 0 \\
0 & 1 & 0 & 0 \\
0 & 0 & 0 & 1 \\
0 & 0 & 1 & 0
\end{bmatrix} = \mathrm{CNOT}_{1\rightarrow 2}.
\end{equation}

For $N$ modes the appropriate transformation is
\begin{equation}
\begin{aligned}
P_{JW\rightarrow P} &= \mathrm{CNOT}_{N-1\rightarrow N}\mathrm{CNOT}_{N-2\rightarrow N-1}\\ &\dots\mathrm{CNOT}_{1\rightarrow 2}.
\end{aligned}
\end{equation}

Via conjugation by $P_{JW\rightarrow P}$, the Jordan-Wigner Majorana operators $\gamma_j$ are transformed to the parity basis Majorana operators $\widetilde{\gamma}_j$,
\begin{equation}
\begin{aligned}
    \widetilde{\gamma}_j  &= P_{JW\rightarrow P} \gamma_j P_{JW\rightarrow P}^\dagger\\
    \widetilde{\gamma}_j' &= P_{JW\rightarrow P} \gamma_j' P_{JW\rightarrow P}^\dagger.
\end{aligned}
\end{equation}

A similar basis permutation can be constructed relating Jordan-Wigner and the Bravyi-Kitaev encoding.

In general, there are $2^N!$ permutations on the computational basis, each of which maps from Jordan-Wigner to a valid fermion-to-qubit encoding with a corresponding set of Majorana operators. It is worth noting that since commutation relations are preserved under conjugation by a unitary operator,
\begin{equation}
\begin{aligned}
    [UAU^\dagger,UBU^\dagger]_{\pm} &= U[A,B]_{\pm}U^\dagger,
\end{aligned}
\end{equation}
the Majorana commutation relations~\eqref{eq:MajoranaCommutationRelations} in particular are preserved under basis permutations. 

\section{A Minimal Qubit Basis}\label{sec:minimalQubitBasis}

Now consider a Jordan-Wigner encoding of an $N$ mode system with $K$ fermions. There are $N \choose K$ possible states; we will find it helpful to index them by the lexicographical order of their associated binary strings. For example, for a system of four modes and two fermions the states are indexed 
\begin{equation}\label{eq:4Mode2FermionExample}
\begin{aligned}
\ket{0011} &\rightarrow 0, \hspace{3mm}
\ket{0101} \rightarrow 1, \hspace{3mm} \ket{0110} \rightarrow 2,\\ \hspace{3mm} \ket{1001} &\rightarrow 3,\hspace{3mm}
\ket{1010} \rightarrow 4,\hspace{3mm} 
\ket{1100} \rightarrow 5.
\end{aligned}
\end{equation}

The main thrust of this work is that it is always possible to construct a permutation $P$ that maps these states
to a basis in which their first $\lceil \log_2 {N \choose K} \rceil$ qubits are sufficient to uniquely specify them. There are in general multiple choices of permutation that accomplish this. Consider for example the permutations which map them to the states whose first $\lceil \log_2 {N \choose K} \rceil$ qubits store their indices as binary strings, and whose remaining qubits are all zeros. (Given that we have freedom to arbitrarily permute the basis states corresponding to other particle number sectors, there are $\left(2^N-{N\choose K}\right)!$ possible permutations that satisfy this requirement.)

Under such a permutation the states in \eqref{eq:4Mode2FermionExample} become
\begin{equation}
\begin{aligned}
P\ket{0011} &= \ket{0000}, \hspace{3mm}
P\ket{0101} = \ket{0010},\\
P\ket{0110} &= \ket{0100}, \hspace{3mm}
P\ket{1001} = \ket{0110},\\ 
P\ket{1010} &= \ket{1000}, \hspace{3mm}
P\ket{1100} = \ket{1010},
\end{aligned}
\end{equation}
where $P$ might be the permutation matrix given in cycle notation~\cite{dummit2003abstract} as 
\begin{equation}
P = (1,4)(3,6)(5,7,10)(9,11,13). 
\end{equation}
Note that this permutation acts on the $16=2^4$ states of the four-qubit system (as opposed to acting on the four qubits directly, e.g.~with the permutations generated by SWAP gates). In general the permutations we consider are elements of $S_{2^N}$ rather than $S_N$.

An alternative choice of $P$ that does not follow the above prescription, but which does nonetheless map to a basis in which the final qubit of all the two-particle states is always zero is
\begin{equation}
P = \mathrm{CNOT}_{1\rightarrow 4}\mathrm{CNOT}_{2\rightarrow 4}\mathrm{CNOT}_{3 \rightarrow 4}.    
\end{equation}

\section{Permutations and the Clifford Group}\label{sec:cliffords}

If a permutation is also an element of the Clifford group, then it is guaranteed by the Gottesman-Knill theorem~\cite{gottesman_heisenberg_1998} that its action on Pauli strings maps them to Pauli strings, rather than linear combinations thereof.

As it is desirable to reduce the number of terms of the encoded Hamiltonian, it is therefore preferable (though not necessary) that the permutation we construct to map from the JW basis to a minimal qubit basis is also a Clifford. This is possible but not guaranteed; for example, the CNOT gate is a permutation matrix and also a Clifford. However, the Toffoli gate is a permutation matrix that is not a Clifford.

In fact, however, we show in the Appendix that the qubit requirement of an encoding corresponding to a Clifford permutation can be reduced by at most 1. Thus, the parity basis encoding is already ``optimal'' if we restrict ourselves to Clifford permutations.



\section{Four Mode Example}\label{sec:example}

In this section, we demonstrate the construction of a minimal qubit basis for the one- and two-fermion sectors of a four-mode system. We note that the three-fermion sector is equivalent to the one-fermion sector by particle-hole symmetry, and the zero- and four-fermion sectors are trivial. We begin with the two-fermion sector, for which a suitable Clifford permutation exists. 

\subsection{Two-fermion sector}

The Clifford permutation
\begin{equation}
P = \mathrm{CNOT}_{1\rightarrow 4}\mathrm{CNOT}_{2\rightarrow 4}\mathrm{CNOT}_{3 \rightarrow 4}    
\end{equation}
acts on the basis states as
\begin{equation}
\begin{aligned}
P\ket{0011} &= \ket{0010}, \hspace{3mm}
P\ket{0101} = \ket{0100},\\
P\ket{0110} &= \ket{0110}, \hspace{3mm}
P\ket{1001} = \ket{1000},\\ 
P\ket{1010} &= \ket{1010}, \hspace{3mm}
P\ket{1100} = \ket{1100},
\end{aligned}
\end{equation}
rendering the final qubit redundant. The Majoranas in this new basis are given by
\begin{equation}
\begin{aligned}
P \gamma_1 P^\dagger &= X_1 I_2 I_3 X_4, &&P\gamma_1' P^\dagger = Y_1 I_2 I_3 X_4,\\
P \gamma_2 P^\dagger &= Z_1 X_2 I_3 X_4, &&P\gamma_2' P^\dagger = Z_1 Y_2 I_3 X_4,\\
P \gamma_3 P^\dagger &= Z_1 Z_2 X_3 X_4, &&P\gamma_3' P^\dagger = Z_1 Z_2 Y_3 X_4,\\
P \gamma_4 P^\dagger &= Z_1 Z_2 Z_3 X_4, &&P\gamma_4' P^\dagger = I_1 I_2 I_3 Y_4,
\end{aligned}
\end{equation}
where the $\gamma_j, \gamma'_j$ are the Jordan-Wigner Majoranas.

Note that number-conserving operators constructed from these Majoranas necessarily act as the identity on the redundant qubits, and we can therefore treat the ``effective'' operator locality of the Majoranas as if they acted only on the first 3 (in general $\lceil \log_2 {N \choose K} \rceil$) qubits.



\subsection{One-fermion sector}

In this case, the qubit requirement can be reduced by two, though no Clifford permutation can accomplish this. A na{\"i}ve choice of non-Clifford permutation, 
\begin{equation}
P_1 = (1,3)(2,13)
\end{equation}
acts on the basis states as
\begin{equation}\label{eq:p1BasisStates}
\begin{aligned}
P_1\ket{1000} &= \ket{1000}, \hspace{3mm} P_1\ket{0100} = \ket{0100},\\
P_1\ket{0010} &= \ket{0000}, \hspace{3mm}
P_1\ket{0001} = \ket{1100},\\
\end{aligned}
\end{equation}
rendering the last two qubits redundant. We omit the resultant Majoranas (of up to 32 terms) for brevity.

It is however possible to find a suitable permutation whose circuit decomposition is Clifford up to a single Toffoli gate,
\begin{equation}
P_2 = \begin{quantikz}
& \gate{X} & \ctrl{2} & \gate{X} & \qw & \targ{} & \ctrl{3} & \qw \\
& \ctrl{1} & \qw & \qw & \targ& \qw & \qw & \ctrl{} & \qw \\
& \targ{} & \targ{} & \targ{} & \qw & \qw & \qw & \qw \\
& \qw & \qw & \ctrl{-1} & \ctrl{-2} & \ctrl{-3} & \targ{} &\qw
\end{quantikz},
\end{equation}
which acts identically to $P_1$ on the basis states \eqref{eq:p1BasisStates}, and leads to transformed Majoranas of up to only 4 terms, e.g.
\begin{equation}
\begin{aligned}
P_2 \gamma_1 P_2^\dagger &= \frac{1}{2}(X_1I_2X_3I_4 + X_1I_2X_3X_4\\ &+ X_1Z_2X_3I_4 - X_1Z_2X_3X_4).
\end{aligned}
\end{equation}

\section{Complexity and Comparison with other qubit reduction schemes}\label{sec:comparison}

In general the number of terms per Majorana scales as $O\left(4^t\right)$, where $t$ is the number of Tofolli gates in the circuit decomposition of the permutation, up to a worst-case of $O\left(4^N\right)$ terms for the encoded Hamiltonian as a whole.

It is not necessary to actually perform a permutation as a quantum circuit to implement the encoding. We note however that our permutations can be decomposed into a product of $O{N \choose K}$ two-level permutations (i.e. operators that swap exactly two states). Using Gray codes~\cite{nielsen_quantum_2010}, such a two-level unitary can be decomposed into a circuit of $O(N^2)$ CNOTs and single-qubit gates. This gives a total circuit complexity of $O\left({N \choose K}N^2\right) = O(N^{K+2})$.

See~\cite{Kirby:2021vkt} for a comparison of the qubit and gate costs of prior second-quantized fermion-to-qubit encodings. Here we note in particular the work of Moll et al.~\cite{moll_optimizing_2016} and Shee et al.~\cite{shee_qubit-efficient_2022}, which also achieve the minimum qubit requirement of $\lceil \log_2 {N \choose K} \rceil$ for particle-number conserving systems by projecting into the subspace of Hilbert space corresponding to the relevant particle number sector. Our work differs in that we present a class of fermion-to-qubit transforms specified by permutation operators. This class of transforms affords the potential for reduced Majorana complexity if permutations whose circuit decompositions contain a minimal number of non-Clifford gates can be found.

\section{Conclusion}\label{sec:conclusion}

In this paper, we have demonstrated that for particle-number conserving systems of fermions, it is possible to construct a permutation operator that maps the Jordan-Wigner encoding to a basis with a provably minimal qubit requirement. This is particularly relevant to the practical implementation of quantum simulation in the present NISQ era of quantum computation.

We also showed that there are in general a large number of choices of permutation that accomplish this mapping. Future work might investigate which choice is optimal. 
Possible applications to local encodings~\cite{chien_custom_2020} and to other symmetries such as spin and point group and translational symmetries are also of interest.

The authors are grateful to Riley Chien for helpful discussions. 
BH and JDW were supported by the US NSF grant PHYS-1820747. JDW was additionally supported by NSF (EPSCoR-1921199) and by the Office of Science, Office of Advanced Scientific Computing Research under programs Fundamental Algorithmic Research for Quantum Computing and Optimization, Verification, and Engineered Reliability of Quantum Computers project. This paper was also supported by the ``Quantum Chemistry for Quantum Computers'' project sponsored by the DOE, Award DE- SC0019374. JDW holds concurrent appointments at Dartmouth College and as an Amazon Visiting Academic. This paper describes work performed at Dartmouth College and is not associated with Amazon.
DA was supported by the U.S. Department
of Energy, Office of Science, Office of Nuclear Physics under Award Number DE-SC0004286.

\bibliographystyle{unsrt}
\bibliography{citations}

\begin{thebibliography}{10}

\bibitem{brod_bosons_2021}
Daniel~Jost Brod.
\newblock Bosons vs. {Fermions} – {A} computational complexity perspective.
\newblock {\em Revista Brasileira de Ensino de Física}, 43, March 2021.

\bibitem{brown_using_2010}
Katherine~L. Brown, William~J. Munro, and Vivien~M. Kendon.
\newblock Using {Quantum} {Computers} for {Quantum} {Simulation}.
\newblock {\em Entropy}, 12(11):2268--2307, November 2010.

\bibitem{feynman_simulating_1982}
Richard~P. Feynman.
\newblock Simulating physics with computers.
\newblock {\em International Journal of Theoretical Physics}, 21(6):467--488,
  June 1982.

\bibitem{lloyd_universal_1996}
Seth Lloyd.
\newblock Universal {Quantum} {Simulators}.
\newblock {\em Science}, 273(5278):1073--1078, August 1996.

\bibitem{Bravyi_2002}
Sergey~B. Bravyi and Alexei~Yu. Kitaev.
\newblock Fermionic {Quantum} {Computation}.
\newblock {\em Annals of Physics}, 298(1):210--226, May 2002.

\bibitem{Jordan:1928wi}
P.~Jordan and E.~Wigner.
\newblock About the {Pauli} exclusion principle.
\newblock {\em Zeitschrift für Physik}, 47(9-10):631--651, September 1928.

\bibitem{doi:10.1063/1.4768229}
Jacob~T. Seeley, Martin~J. Richard, and Peter~J. Love.
\newblock The {Bravyi}-{Kitaev} transformation for quantum computation of
  electronic structure.
\newblock {\em The Journal of Chemical Physics}, 137(22):224109, December 2012.
\newblock arXiv: 1208.5986.

\bibitem{havlicek_operator_2017}
Vojtěch Havlíček, Matthias Troyer, and James~D. Whitfield.
\newblock Operator locality in the quantum simulation of fermionic models.
\newblock {\em Physical Review A}, 95(3):032332, March 2017.

\bibitem{kassal2008polynomial}
Ivan Kassal, Stephen~P Jordan, Peter~J Love, Masoud Mohseni, and Al{\'a}n
  Aspuru-Guzik.
\newblock Polynomial-time quantum algorithm for the simulation of chemical
  dynamics.
\newblock {\em Proceedings of the National Academy of Sciences},
  105(48):18681--18686, 2008.

\bibitem{zalka1998simulating}
Christof Zalka.
\newblock Simulating quantum systems on a quantum computer.
\newblock {\em Proceedings of the Royal Society of London. Series A:
  Mathematical, Physical and Engineering Sciences}, 454(1969):313--322, 1998.

\bibitem{Kirby:2021vkt}
William Kirby, Bryce Fuller, Charles Hadfield, and Antonio Mezzacapo.
\newblock Second-{Quantized} {Fermionic} {Operators} with {Polylogarithmic}
  {Qubit} and {Gate} {Complexity}.
\newblock {\em PRX Quantum}, 3(2):020351, June 2022.

\bibitem{dummit2003abstract}
D.S. Dummit and R.M. Foote.
\newblock {\em Abstract Algebra}.
\newblock Wiley, 2003.

\bibitem{gottesman_heisenberg_1998}
Daniel Gottesman.
\newblock The {Heisenberg} {Representation} of {Quantum} {Computers}.
\newblock July 1998.
\newblock arXiv: quant-ph/9807006.

\bibitem{nielsen_quantum_2010}
Michael~A. Nielsen and Isaac~L. Chuang.
\newblock {\em Quantum computation and quantum information}.
\newblock Cambridge University Press, Cambridge ; New York, 10th anniversary ed
  edition, 2010.

\bibitem{moll_optimizing_2016}
Nikolaj Moll, Andreas Fuhrer, Peter Staar, and Ivano Tavernelli.
\newblock Optimizing qubit resources for quantum chemistry simulations in
  second quantization on a quantum computer.
\newblock {\em Journal of Physics A: Mathematical and Theoretical},
  49(29):295301, July 2016.

\bibitem{shee_qubit-efficient_2022}
Yu~Shee, Pei-Kai Tsai, Cheng-Lin Hong, Hao-Chung Cheng, and Hsi-Sheng Goan.
\newblock Qubit-efficient encoding scheme for quantum simulations of electronic
  structure.
\newblock {\em Physical Review Research}, 4(2):023154, May 2022.

\bibitem{chien_custom_2020}
Riley~W. Chien and James~D. Whitfield.
\newblock Custom fermionic codes for quantum simulation.
\newblock September 2020.
\newblock arXiv: 2009.11860 [quant-ph].

\bibitem{bravyi_hadamard-free_2021}
Sergey Bravyi and Dmitri Maslov.
\newblock Hadamard-{Free} {Circuits} {Expose} the {Structure} of the {Clifford}
  {Group}.
\newblock {\em IEEE Transactions on Information Theory}, 67(7):4546--4563, July
  2021.

\bibitem{bataille_quantum_2020}
Marc Bataille.
\newblock Quantum circuits of {CNOT} gates, December 2020.
\newblock arXiv:2009.13247 [quant-ph].

\end{thebibliography}

\section{Appendix: Clifford Permutations}

The set of Cliffords that map basis vectors to basis vectors (up to a possible phase) is generated by the gates \{CNOT, CZ, X, P\} ~\cite{bravyi_hadamard-free_2021}. Of these, CZ and P only modify phases, and thus the set of Clifford permutations is generated by CNOT's and X's.

Now, due to the following identities,

\begin{equation}
\begin{aligned}
    \mathrm{CNOT}_{1\rightarrow 2}X_1 &= X_1X_2 \mathrm{CNOT}_{1\rightarrow 2}\\
    \mathrm{CNOT}_{1\rightarrow 2}X_2 &= X_2 \mathrm{CNOT}_{1\rightarrow 2}
\end{aligned}
\end{equation}

we see that a circuit composed of CNOT's and X's can be reduced to a circuit composed of only CNOTs, with all the X's moved to one side. Since such a layer of X's cannot change the qubit redundancy properties of our states of interest, we need only consider the set of $S_{2^n}$ permutation matrices that can be generated by CNOTs.

Now, the group generated by CNOTs on $n$ qubits is isomorphic to $\mathrm{GL}_n(\mathbb{F}_2)$~\cite{bataille_quantum_2020}, the set of $n \times n$ invertible matrices over the finite field with two elements (i.e.~0 and 1). This allows us to represent unitaries generated by CNOTs as $n \times n$ matrices that act directly on the binary string representations of the qubit states, e.g. the equation $\mathrm{CNOT}\ket{10} = \ket{11}$ can  be represented using the $\mathrm{GL}_n(\mathbb{F}_2)$ isomorphism by
\begin{equation}
\begin{bmatrix}
1 & 0 \\
1 & 1
\end{bmatrix}\begin{bmatrix} 1\\0 \end{bmatrix} = \begin{bmatrix} 1\\1 \end{bmatrix},    
\end{equation}
where it is understood that addition is modulo 2.

Now consider a $2^n \times 2^n$ permutation operator acting on a set of qubit states whose binary string representations all have identical Hamming weight $k$. If this operator maps these strings to strings with a digit that is the same for all strings, then the corresponding qubit is redundant, as we have discussed. We have seen that the parity mapping accomplishes precisely this with a series of CNOTs, rendering the last qubit in the encoding redundant in the presence of parity conservation. We will now show that no series of CNOTs (and hence no Clifford operator) can render more than one qubit redundant in this way.

\begin{theorem}
A matrix $M \in \mathrm{GL}_n(\mathbb{F}_2)$ acting on the set of binary strings of length $n$ and constant Hamming weight $0 < k < n$ cannot map this set to a set of strings across which two or more digits are identical. 
\end{theorem}

\begin{proof}
Suppose we have some matrix $M \in \mathrm{GL}_n(\mathbb{F}_2)$ that maps all binary strings of length $n$ with Hamming weight $0 < k < n$ to strings whose last two (or more) digits are identical. We need consider only the last digits since CNOTs generate SWAP gates and we can therefore permute the digits arbitrarily anyway.

Accordingly, consider the last two rows of the matrix. We need them to have the property that their dot product with any binary string of length $n$ and Hamming weight $k$ gives the same digit - either always 1 or always 0. 

Since this matrix is invertible, it can have at most one row consisting entirely of 1s, and no rows consisting entirely of 0s. Therefore at least one of the last two rows must have at least one 1 and at least one 0. Without loss of generality, assume that the row begins $(1,0, \dots)$. Now consider two binary strings with equal Hamming weight that begin $(1, 0, \dots)$ and $(0, 1, \dots)$ and are identical after the first two digits. The row will have a different dot product with each of these, a contradiction.
\end{proof}

Essentially, every bit of the output string of the mapping is the XOR of a non-empty subset of the input string. But the only such subset that produces the same result for every input string of Hamming weight $k$ is the entire set.    

\end{document}